\documentclass[pre,twocolumn,superscriptaddress,showpacs,amsmath,amssymb]{revtex4}
\usepackage{graphicx,epstopdf,dcolumn,bm} 
\usepackage{euscript,graphicx,psfrag,amscd,amsgen,amsfonts,amssymb,latexsym}

\begin{document}

\title{Time$-$metric equivalence and dimension change under time reparameterizations}
\author{Adilson E. Motter}
\affiliation{Department of Physics and Astronomy, Northwestern University, Evanston, IL 60208, USA}
\affiliation{Northwestern Institute on Complex Systems, Northwestern University, Evanston, IL 60208, USA}
\author{Katrin Gelfert}
\affiliation{Department of Mathematics, Northwestern University, Evanston, IL 60208, USA}

\date{\today}

\begin{abstract}

We study the behavior of dynamical systems under time reparameterizations,  which is
important not only to characterize chaos in relativistic systems but also to probe the 
invariance of dynamical quantities. We first show that 
time transformations are locally equivalent to metric transformations, a result
that leads to a transformation rule for all Lyapunov exponents on arbitrary Riemannian phase spaces. 
We then show that time transformations preserve the spectrum of generalized dimensions $D_{q}$ except
for the information dimension $D_1$, which, interestingly, transforms in a nontrivial way despite previous
assertions of invariance. 
The discontinuous behavior at $q=1$
can be used to constrain and extend the formulation
of the Kaplan-Yorke conjecture.
\end{abstract}
\pacs{05.45.-a, 04.20.Cv}

\maketitle

Recent studies of chaos in general relativity and cosmology highlighted the importance of 
the time parameterization as an extra dimension in the characterization of chaotic
dynamics~\cite{Mot:03}.
Ever since Lyapunov exponents and other dynamical quantities were found to depend on the choice 
of the time parameter~\cite{FraMat:88}, much effort has been directed towards an invariant characterization 
of chaos that would {\it avoid} the difficulties imposed by this dependence~\cite{Hobill:94,Gurzadyan:00,Cornish:97,Cipriani:98,Mot:2001}. 
However, at the most fundamental level, one could instead seek to {\it explore} the freedom introduced by time 
transformations in order to investigate the dependence of the dynamical quantities on the geometrical versus temporal
properties of the orbits, which is an important and often elusive open problem.
This problem can be traced to the  question of {\it how} dynamical quantities
change under time transformations.

In this Rapid Communication, we consider the dynamical effect of spatially  
inhomogeneous time transformations (i.e., time reparameterizations that depend on 
the phase-space coordinates).
From the perspective of 
the rate of
separation between nearby trajectories, we 
show that the 
time reparameterizations can be identified with local transformations of the phase-space 
metric. This implies that all Lyapunov exponents of a given orbit
are scaled by a common factor and the resulting Lyapunov dimension is invariant under time transformations. 
We show, however, that the information dimension is generally not invariant in non-ergodic systems, 
illustrating that the identity between the information dimension and the Lyapunov dimension of average Lyapunov exponents 
generally does not hold in such systems;
noticeably, the other generalized dimensions remain invariant in spite of their dependence on the invariant measure, 
which does change.

Numerous physical systems can be described as smooth dynamical systems of the form
\begin{equation}\label{eq1}
\frac{d\boldsymbol x}{dt} = \boldsymbol f(\boldsymbol x)
\end{equation} 
defined on a smooth Riemannian manifold $M$ of certain metric $g$, which represents the 
phase space of the 
system. 
We focus on this general class of systems and
consider time reparameterizations of the form
\begin{equation}\label{eq1t}
d\tau = r(\boldsymbol x) dt,
\end{equation}  
where $r$ is a smooth and strictly positive integrable function. 
We also assume that $r$ and $r^{-1}$ are bounded away from zero
on the asymptotic sets of the system.
These conditions assure that $\tau(\boldsymbol x_o,t)=\int_{0}^t r(\boldsymbol x(t'))dt'$, 
where $\boldsymbol x_o\equiv \boldsymbol x(0)$, is a well-defined time parameter. 
In the case of Friedmann-Robertson-Walker 
cosmological models, for instance, the proper 
time $T$ and the conformal time $\eta$ are related through the relation $dT=ad\eta$,
where the dynamical variable $a$ is positive away from cosmological singularities~\cite{PRD2002}.
But the reparameterization (\ref{eq1t}) is not limited to relativistic systems, in that it can represent 
any change of independent variable; parameter $\tau$ could be, for example, a monotonically increasing 
angular coordinate.

We first note~\cite{CorFomSin:82} that the time reparameterization changes an
invariant probability measure from $\mu$ to $\mu_r$ according to   
\begin{equation}\label{sua}
d\mu_ r = \frac{ r}{\int_M r d\mu} d\mu \,.
\end{equation}
This change applies, in particular, to natural probability measures,   
despite the fact that the orbits remain invariant and ergodicity 
is preserved by time reparameterizations~\cite{CorFomSin:82}.
Physically, this reflects the fact that
the transformed system evolves at different 
speeds and hence with different residence times along the orbits~\cite{result0}. 

Next we note that the Lyapunov exponents,                 
\begin{equation}\label{deflya}
  \lambda(\boldsymbol x_o,\boldsymbol v_o) =
  \limsup_{t\to\infty}\frac{1}{t}\log\lVert \boldsymbol v(\boldsymbol x_o,t)\rVert,
\end{equation}
may change as the time is reparameterized~\cite{FraMat:88,Mot:03}.
Here $\boldsymbol v(\boldsymbol x_o,t)$ is the solution of the variational equation of system
\eqref{eq1}
for an initial condition $\boldsymbol x_o$ and an initial vector $\boldsymbol v_o$ modeling the distance
between nearby trajectories, and $\lVert \cdot \rVert$ is the norm induced by 
the Riemannian metric. The time transformation generally changes the length
and direction of the vectors $\boldsymbol v(\boldsymbol x_o,t)$ for $t>0$, 
which are then denoted by $\boldsymbol v_r(\boldsymbol x_o,\tau(\boldsymbol x_o,t))$.
However, we now show that an equivalent change can be induced by a transformation
of the metric.

Specifically, we construct a  Riemannian metric $\widetilde g$ on $M$ such that 
for any two vectors $\boldsymbol v_o$ and $\boldsymbol w_o$ in the tangent space $T_{\boldsymbol x_o}M$ 
of $M$ at $\boldsymbol x_o$ we have
\begin{equation}
\left \langle \boldsymbol v(\boldsymbol x_o,t),\boldsymbol w(\boldsymbol x_o,t)\right \rangle_{\widetilde g} = 
\left \langle \boldsymbol v_r(\boldsymbol x_o,\tau(\boldsymbol x_o,t)),\boldsymbol w_r(\boldsymbol x_o,\tau(\boldsymbol x_o,t))\right \rangle_{g}
\label{neq24}
\end{equation}
for every $t$ in some interval $I$ around zero. 
Here $\boldsymbol v$ and $\boldsymbol w$ ($\boldsymbol v_r$ and $\boldsymbol w_r$)
correspond to the
solutions of the variational equation before (after) the time reparameterization,
and $\langle\cdot,\cdot\rangle$ stands for the scalar product induced by the metric. We
say that $\widetilde g$ and $g$ satisfying \eqref{neq24} are  \emph{locally related} by the time 
reparameterization $r$ at $\boldsymbol x_o$.

To proceed we consider $N$=\,$\dim M$ linearly independent vectors $\boldsymbol v_j\in T_{\boldsymbol  x_o}M$, $\lVert
\boldsymbol v_j\rVert=1$,  and denote by $\boldsymbol y_j(t)\equiv \boldsymbol y(\boldsymbol x_o,\boldsymbol v_j,t)$
the solution of the variational equation of \eqref{eq1} with $\boldsymbol y_j(0)=\boldsymbol v_j$.
With respect to a local representation
(and using the summation convention), the variational equation reads
\begin{equation}
\frac{d{y_j}^k}{dt} = \frac{\partial f^k}{\partial x^i}{y_j}^i, \;\; k=1, \ldots, N,
\end{equation}
and {\it does not} depend on the metric. The same argument applies to 
the solutions $\boldsymbol z_j(t)\equiv \boldsymbol z(\boldsymbol x_o,\boldsymbol v_j,\tau(\boldsymbol x_o,t))$ of 
the variational equation after the time transformation, 
\begin{equation}
\frac{d{z_j}^k}{d\tau} = \frac{\partial f^k/r}{\partial x^i}{z_j}^i, \;\; k=1, \ldots, N,
\end{equation}
with $\boldsymbol z_j(0)=\boldsymbol v_j$. 

On account of this, condition~\eqref{neq24} can be restated as 
\begin{equation}\label{equas}
\langle \boldsymbol y_i (t),\boldsymbol y_j(t)\rangle_{\widetilde g} = 
\langle \boldsymbol z_i (t),\boldsymbol z_j (t)\rangle_{g} 
\end{equation}
for every $i$, $j=1, \ldots, N$ and every $t\in I$. 
Representing the metric tensor $g$ locally by 
$G(\boldsymbol x_o,t)\equiv\left( g_{ij}(\boldsymbol x(t))\right)$,
Eqs.~\eqref{equas} determine the choice of a family of matrices 
$G(\boldsymbol x_o,t)$ along the trajectory $\boldsymbol x(t)$. 
The corresponding system of $N^2$ equations can be written in matrix
form as 
\begin{equation}\label{finali}
{Y_t}^{\dagger} \widetilde G(\boldsymbol x_o,t) Y_t = {Z_t}^{\dagger} G(\boldsymbol x_o,t) Z_t,
\end{equation}
where $^\dagger$ denotes the matrix transpose,
$\widetilde G \equiv \left( \widetilde g_{ij} \right)$ denotes the local representation
of the metric tensor $\widetilde g$, 
and where
$Y_t\equiv(\boldsymbol y_1(t),\boldsymbol y_2(t),\ldots,\boldsymbol y_N(t))$ and
$Z_t\equiv(\boldsymbol z_1(t),\boldsymbol z_2(t),\ldots,\boldsymbol z_N(t))$ 
are $N\times N$-matrices having the vectors 
${\bm y}_j(t)$ and $\bm z_j(t)$ as columns.
The matrix $Y(t)$ is invertible since $\boldsymbol v_1, \ldots, \boldsymbol v_N$ are linearly 
independent and the linear map $\boldsymbol v_o\rightarrow \boldsymbol v(\boldsymbol x_o,t)$ is invertible.  
The latter follows from the fact that the flow map $\boldsymbol x_o\rightarrow \boldsymbol x(t)$ is a 
diffeomorphism on $M$. Therefore, \eqref{finali} leads to
\begin{equation}\label{metric}
\widetilde G(\boldsymbol x_o,t) = (Z_t {Y_t}^{-1})^\dagger G(\boldsymbol x_o,t) Z_t {Y_t}^{-1},
\end{equation}
which is the locally related metric that we sought to construct.

    \begin{figure}[h]
      \includegraphics[width=0.9\linewidth]{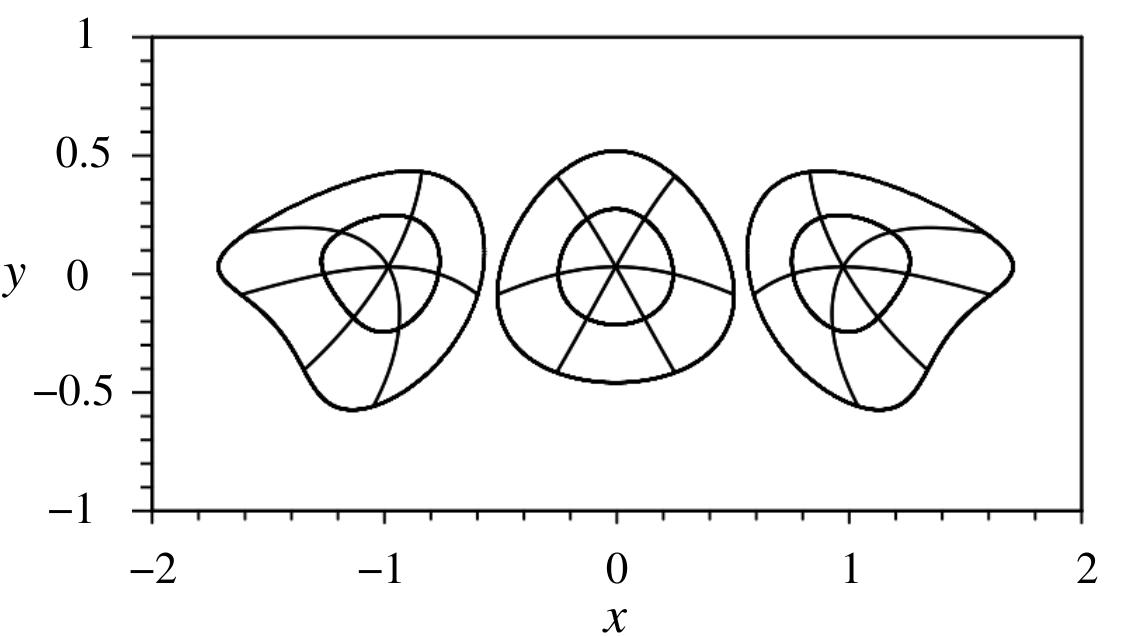}  
      \caption{Time-metric equivalence. Geodesic coordinates of metric~\eqref{example} for three different initial 
points and contour lines 
 at distances $1/4$ and $1/2$ from these points. In this illustration, all geodesic lines 
have length 1/2 with respect to the new metric.
      } 
    \label{fig1}
    \end{figure}

A simple example is illustrated 
in Fig.~\ref{fig1}, in which we time transform
the linear flow $dx/dt = 1$, $dy/dt =0$ with $r(x,y)=e^{-y}$. 
Given $x(0)=0$ and arbitrary $y(0)$, the metric $\widetilde g$
that is locally related to the 2D Euclidean metric is given by the 
matrix
\begin{equation}\label{example}
 \left(\widetilde g_{ij}\right) = 
\left(\begin{matrix}1&x\\x&1+x^2\end{matrix}\right).
\end{equation}
In the figure we show local geodesic coordinates of the new metric, that is, given an initial
point we draw the outgoing geodesics of fixed lengths with respect to $\widetilde g$. 
Note that even this simple system
exhibits interesting properties due to the shear introduced by the time reparameterization. 
A few observations are in order.

First, the metric $\widetilde g$ defined by \eqref{metric} does not depend on the
initial choice of vectors $\boldsymbol v_j$, $j=1,\ldots, N$. Because 
the variational equations are linear, we can write $Y_t=\Phi(\boldsymbol x_o,t) V_0$ 
and $Z_t= \Phi_r(\boldsymbol x_o,\tau(\boldsymbol x_o,t)) V_0$, 
where $V_0=(\boldsymbol v_1,\boldsymbol v_2,\ldots,\boldsymbol v_N)$
is the $N\times N$-matrix having the vectors $\boldsymbol v_j$ 
as columns and where $\Phi$ and $\Phi_r$ are the local representations 
of the evolution matrices. This leads to
$\widetilde G(\boldsymbol x_o,t) = W(\boldsymbol x_o,t)^\dagger G(\boldsymbol x_o,t) W(\boldsymbol x_o,t)$, 
where $W(\boldsymbol x_o,t)\equiv \Phi_r(\boldsymbol x_o,\tau(\boldsymbol x_o,t)) \Phi ^{-1}(\boldsymbol x_o,t)$.
Second, the metric $\widetilde g$ depends smoothly on the initial conditions and
can be extended in a neighborhood of $\boldsymbol x_o$; 
a smooth extension can be generated for every given smooth surface of initial points 
passing through 
$\boldsymbol x_o$ transversely to the flow.
Third, the dependence of the metric $\widetilde g$ on the time $t$ indicates that,
in general, the metric is
single-valued 
\emph{only} on some finite time 
interval along the trajectories. For example, the interval $I$ 
is generally limited by $t_0$ if  
$\boldsymbol x_o$ is a periodic point of period $t_0$ 
(with respect to the time $t$) and the function $r$ is such that 
$\Phi(\boldsymbol x_o,t_0)\neq\Phi_r(\boldsymbol x_o,\tau(\boldsymbol x_o,t_0))$.
The interval $I$ is similarly constrained
by the recurrence of orbits 
to the neighborhood in which the metric is extended~\cite{local_global}. 
Furthermore, the time dependence of the metric can 
be eliminated in favor of a dependence on $\boldsymbol x_o$ and $\boldsymbol x$ only.  
Therefore, our result establishes a {\it local} equivalence between time and metric
transformations on $M$.

A neat implication of this time-metric equivalence is that, under the time
reparameterization~\eqref{eq1t}, the Lyapunov exponent in Eq.~\eqref{deflya}
changes exclusively due to the
transformation of the factor $1/t$.
The contribution due to 
the logarithmic factor remains unchanged because the norms induced
by the metrics $g$ and  $\widetilde g$ are 
logarithmically equivalent 
along each orbit \cite{local_global};     
that is, 
for each ${\boldsymbol x_o}$ and ${\boldsymbol v_o}$ there is a
sub-exponential function $C=C(t)$
such that
$ ||\boldsymbol v(\boldsymbol x_o, t)||_{\widetilde g}
\equiv||W(\boldsymbol x_o,t) \boldsymbol v(\boldsymbol x_o, t)||_g = C(t)||\boldsymbol v(\boldsymbol x_o, t)||_g$
\cite{longpaper}.
Therefore, the Lyapunov exponents transform as 
$\lambda_r(\boldsymbol x_o,\boldsymbol v_o) = \lambda(\boldsymbol x_o,\boldsymbol v_o)/\Lambda(\boldsymbol x_o)$, 
where $\Lambda(\boldsymbol x_o)=\lim_{t\rightarrow \infty} \tau(\boldsymbol x_o,t)/t$. This extends the
result previously derived in Ref.~\cite{Mot:03} for Euclidean phase spaces to the more general case of 
Riemannian manifolds. 

We now turn to the transformations of fractal dimensions, which cannot be accounted for by 
metric changes. The box-counting dimension $D_0$ is purely geometrical and hence does not
change under time reparameterization. The generalized dimensions $ D_q$, however, can in 
principle change for $q>0$ given that they depend on the
measure and the measure 
is transformed according to (\ref{sua}). 
To analyze this dependence, we consider a positive-measure set of interest $S$ 
(typically an attractor) and define the spectrum of dimensions on $S$ as 
\begin{eqnarray}\label{neq49}
D_q(\mu) &=& \frac{1}{q-1}\limsup_{\varepsilon\to 0}\frac{1}{\log\varepsilon} \log\sum_{k=1}^{N(\varepsilon)} \mu(B_k)^q, \; q\ge 0,q\neq 1, \\ 
\label{info}
D_1(\mu) &=& \limsup_{\varepsilon\to 0}\frac{1}{\log\varepsilon} \sum_{k=1}^{N(\varepsilon)} \mu(B_k) \log \mu(B_k),
\end{eqnarray}
where the sum is taken over the $N(\varepsilon)$ nonzero measure boxes $B_k$ of edge length  
$\varepsilon$ necessary to cover the set \cite{grass1983,hents1983,comment}. This spectrum
includes as special cases the information dimension ($q=1$) and the correlation dimension ($q=2$).
For consistency, the measure is always normalized to 1 on $S$ (with $M$ replaced by $S$ in Eq.~\eqref{sua}).
The dimensions $D_q$ are known not to depend on smooth transformations of the phase space. We now 
consider their behavior under time transformations.

The first surprise is that, contrary to what intuition may suggest, the dimensions
defined by Eq.~\eqref{neq49} are invariant with respect to time transformations not only
for $q=0$ but also for all $q\neq 1$. This follows from the fact that $\mu$ and  
$\mu_ r$ in Eq. \eqref{sua} are absolutely continuous with respect to each other, i.e., 
both measures define the same sets of nonzero measure, and that $r$ and $r^{-1}$ are bounded 
away from zero on these sets. 
Then there exist positive constants $c_1$, $c_2$ such that $c_1\mu(B_k)\le \mu_ r(B_k)\le c_2\mu(B_k)$ 
for every $k$ and $\varepsilon>0$.    
Using this in the definition \eqref{neq49}, we obtain
\begin{equation}
D_q(\mu_r) = D_q(\mu)  \;\; \mbox{for } q\ge 0, q\neq 1,
\end{equation}
i.e., the dimensions remain unchanged despite their dependence on the measure, 
which generally changes.

The second surprise is that the information dimension \eqref{info} exhibits a
distinctive behavior and may change under the same time reparameterization. 
To appreciate this, we first notice that Eq.~\eqref{info} can be written in 
the continuous form $D_1(\mu) = \limsup_{\varepsilon\to 0} \frac{1}{\log\varepsilon}
\int_{S}\log\mu(B(\boldsymbol x,\varepsilon))\,d\mu(\boldsymbol x)$, 
where $B(\boldsymbol x,\varepsilon)$ is an open ball of radius $\varepsilon$ centered at $\boldsymbol x$.
We then use the  Fatou lemma to obtain 
\begin{equation}\label{pow}
D_1(\mu) \le \int_S \limsup_{\varepsilon\to 0} \frac{1}{\log\varepsilon}\log\mu(B(\boldsymbol x,\varepsilon))\,d\mu(\boldsymbol x),
\end{equation}
where the integrand is the pointwise dimension, which we denote by $\overline {\cal D}_{\mu}(x)$. A similar inequality
holds for the infimum, in which case the pointwise dimension is denoted by $\underline {\cal D}_{\mu}(\boldsymbol x)$. 
This leads to
\begin{equation}
\int_S \underline {\cal D}_{\mu}(\boldsymbol x) \,d\mu(\boldsymbol x) \le D_1(\mu) \le \int_S  \overline {\cal D}_{\mu}(\boldsymbol x)\,d\mu(\boldsymbol x).
\label{swch}
\end{equation}
In the remaining part of the paper we limit the discussion to the case  $\underline {\cal D}_{\mu}(\boldsymbol x) = \overline {\cal D}_{\mu}(\boldsymbol x)\equiv {\cal D}_{\mu}(\boldsymbol x)$ almost everywhere, a property found in many physical systems and demonstrated for flows with strong hyperbolic behavior \cite{hyper}. 
This assures that the equalities hold in (\ref{swch}).

The transformed information dimension is then written as
  \begin{equation}\label{hicks}
  D_1(\mu_ r) = \int_S\frac{ r(\boldsymbol x)}{\int_S r d\mu}{\cal D}_\mu(\boldsymbol x)d\mu(\boldsymbol x),
  \end{equation} 
where we have used the measure \eqref{sua} normalized on $S$ and the invariance of the pointwise dimension. 
The latter follows from \eqref{sua} and is stated as ${\cal D}_{\mu_r}(\boldsymbol x)= {\cal D}_{\mu}(\boldsymbol x)$ for almost every $\boldsymbol x$.
The result in \eqref{hicks}  indicates that $D_1$ is in general noninvariant when ${\cal D}_\mu(\boldsymbol x)$ is not
almost everywhere constant.

For example, consider a system with two ergodic components, 
$S_A$ and $S_B$,
of information dimension $D_1(\mu|S_A)>D_1(\mu|S_B)$. For simplicity, assume that the original measure is
evenly split between the two sets, i.e. $\mu(S_A)=\mu(S_B)$, and that $\mu(B_k)$ is the same for all the 
nonzero measure boxes $B_k$ of each set. The information dimension of 
$S=S_A\cup S_B$ is $D_1(\mu)=\frac{1}{2}[D_0(S_A)+D_0(S_B)]$.
Now, imagine a time reparameterization that changes $\mu|S_1$ uniformly by a factor $0<\alpha<2$ and  
$\mu|S_2$ uniformly by a factor $\beta=2-\alpha$. The transformed information dimension is
$D_1(\mu_r)=\frac{\alpha}{2}D_0(S_A)+\frac{2-\alpha}{2}D_0(S_B)$, which differs from $D_1(\mu)$ 
for any $\alpha\neq 1$ \cite{entangled}.  

\begin{figure}[t]
      \centering
      \includegraphics[width=0.85\linewidth]{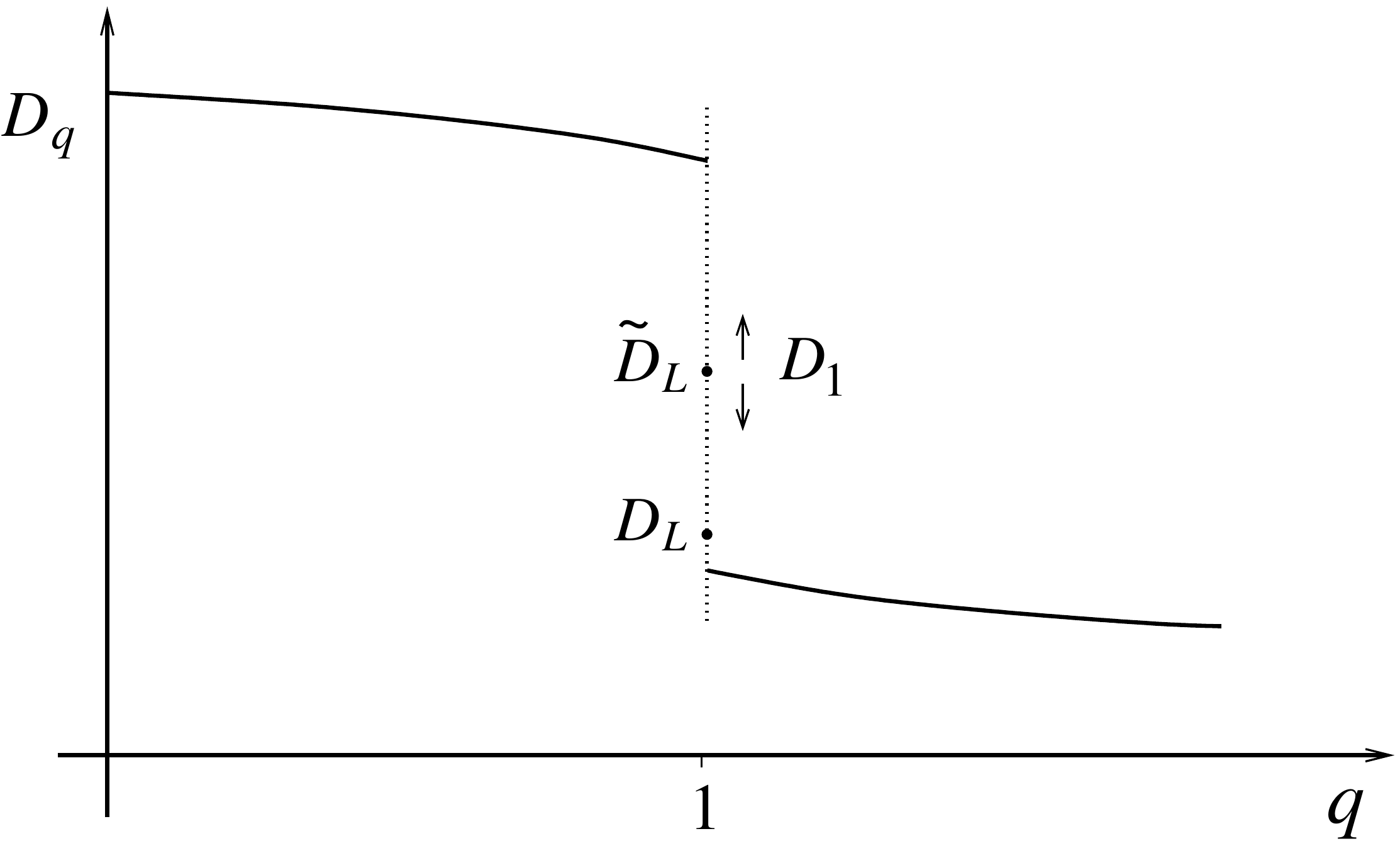}
      \caption{Dimension change. The gap between the generalized dimensions on the left and right 
       sides of $q=1$ is invariant and determines the interval of possible changes for $D_1$ under 
       time transformations.
       The Lyapunov dimension $D_L$ remains invariant, whereas the Lyapunov dimension 
         $\tilde{D}_L$ transforms as $D_1$}.
       \label{fig2} 
    \end{figure}

In the case of $q\neq 1$, this change in the measure contributes an additive term to
$\log \sum_{k} \mu(B_k)^q$ in Eq.~\eqref{neq49} that vanishes when divided by $\log\varepsilon$ 
in the limit of small $\varepsilon$, in agreement with our prediction that the other dimensions 
$D_q$ are all invariant. At first sight the noninvariance for $q=1$ may seem to violate the 
monotonicity of $D_q$, which was previously proved
to hold for $0\le q <1$ and for $q>1$ \cite{grass1983}, but this intuition is misleading 
because $\lim_{q\rightarrow 1-} D_q >\lim_{q\rightarrow 1+} D_q$ 
whenever ${\cal D}_\mu(\boldsymbol x)$ is not constant and $D_1$ is not invariant. 
In our example, when $q<1$, the contribution $\sum_{k} \mu(B_k)^q$ from the set with the largest 
box-counting 
dimension dominates and leads to $D_{q<1}=D_0(S_A)$, just as in the case $q=0$; when $q>1$,
the box-counting dimension of the other set dominates, and $D_{q>1}=D_0(S_B)$. 
The information dimension $D_1(\mu_r)$ is thus a weighted average of the dimensions on both sides
of the discontinuity and is in general free to vary between $\lim_{q\rightarrow 1-} D_q$ and
$\lim_{q\rightarrow 1+} D_q$ under time reparameterizations, as shown in Fig. \ref{fig2}. 

The information dimension is guaranteed to be invariant only in special cases. 
The most important such case is when $\mu$ (and hence $\mu_r$) is ergodic in $S$. 
Since the flow map $\boldsymbol x_o\rightarrow \boldsymbol x(t)$ 
is a diffeomorphism and the measure is invariant under this map,
one can verify that  ${\cal D}_\mu(\boldsymbol x_o)={\cal D}_\mu(\boldsymbol x(t))$.
Then, if $\mu$ is ergodic, ${\cal D}_\mu(\boldsymbol x)$ is 
constant for almost every $\boldsymbol x$, and hence $D_1(\mu_r)=D_1(\mu)$. 
The general condition for $D_1$ to be invariant with respect to any time 
transformation is that ${\cal D}_\mu(\boldsymbol x)$ is constant almost 
everywhere. 

It is of interest to analyze the meaning of the noninvariance of $D_1$ for the Kaplan-Yorke conjecture 
\cite{ld1,ld2} and its generalizations, which state that the information dimension typically equals
the Lyapunov dimension.
Let the average Lyapunov exponents be
$\lambda_i=\int \lambda_i(\boldsymbol x) d\mu(\boldsymbol x)$, where  
$\lambda_1(\boldsymbol x)\ge\lambda_2(\boldsymbol x)\ge\dots  \ge\lambda_N(\boldsymbol x)$ corresponds
to the ordered set of Lyapunov exponents (\ref{deflya}) at $\boldsymbol x$, and assume that
this definition is applied to measures that are not necessarily proved to be ergodic.
Based on the definition operationally used in numerical experiments, 
the Lyapunov dimension can be defined as
\begin{equation}
D_L(\mu)=K+\frac{1}{|\lambda_{K+1}|}\sum_{j=1}^{K}\lambda_i,
\label{dl}
\end{equation} 
where $K$ is the largest integer such that $\sum_{j=1}^{K}\lambda_i\ge 0$,
under the condition that the r.h.s. terms in (\ref{dl}) are well defined.
It follows 
that $D_L(\mu)$ remains invariant under time reparameterizations, 
thus violating the equality $D_L=D_1$ when $D_1$ changes.  In the example considered
above, the Lyapunov dimension of $S=S_A\cup S_B$ is intermediate between the Lyapunov
dimensions of $S_A$ and $S_B$,
indicating that $D_L$ equals $D_1$ for at most one value of $\alpha$. 
The conjecture can be re-established,
however, for the Lyapunov dimension
defined as
\begin{equation}
\tilde{D}_L(\mu)=\int\{ K(\boldsymbol x)+\frac{1}{|\lambda_{K(\boldsymbol x)+1}(\boldsymbol x)|}\sum_{j=1}^{K(\boldsymbol x)}\lambda_i(\boldsymbol x)\} d\mu(\boldsymbol x),
\label{dl2}
\end{equation}
where $K$ is defined as above but now at each point $\boldsymbol x$ (with the convention
that the integrand is zero for $\lambda_1(\boldsymbol x)<0$ and $N$ for $\lambda_N(\boldsymbol x)>0$). It follows 
from (\ref{hicks}) and the Kaplan-Yorke conjecture for typical ergodic sets that
the identity $D_1=\tilde{D}_L$ is expected to hold true for generic systems (see Fig. \ref{fig2}).

The noninvariance of the information dimension, which was previously surmised {\it to be} 
invariant \cite{Cornish:97}, is important as it limits the applicability of the identity
$D_L=D_1$ in nonergodic systems and ergodicity is a property often 
difficult to verify.
The invariance of other indicators of chaos established in this paper 
is
relevant to the study of a range of dynamical phenomena, including 
spatiotemporal chaos,
and clarifies longstanding problems in relativistic chaos.
It shows, in particular, the observer
invariance of the often questioned chaoticity of the mixmaster model for the early universe \cite{Hobill:94}, 
which was first recognized as a chaotic geodesic flow on a Riemannian manifold by Chitre in 1972 in a work that 
made one of the very first uses of the term ``chaos'' in dynamics \cite{chitre72}.


\begin{thebibliography}{99} 

\bibitem{Mot:03}
A.~E.~Motter, Phys. Rev. Lett. \textbf{91}, 231101 (2003); 
A.~E.~Motter and A.~Saa, {\it ibid.} {\bf 102}, 184101 (2009). 

\bibitem{FraMat:88} G.~Francisco and G.~E.~A.~Matsas, Gen. Relativ. Gravit. \textbf{20}, 1047 (1988).

\bibitem{Hobill:94}
D.~Hobill, A.~Burd, and A.~Coley (Eds.), {\it Deterministic Chaos in General Relativity}  (Plenum, 1994).

\bibitem{Gurzadyan:00}
V.~G.~Gurzadyan and R.~Ruffini (Eds.), {\it The Chaotic Universe} (World Scientific, 2000).
   
\bibitem{Cornish:97}
N.~J.~Cornish and J.~J.~Levin, Phys. Rev. D \textbf{55}, 7489 (1997).

\bibitem{Cipriani:98}
P.~Cipriani and M.~Di Bari, Phys. Rev. Lett. \textbf{81}, 5532 (1998).

\bibitem{Mot:2001}
A.~E.~Motter and P.~S.~Letelier, Phys. Lett. A \textbf{285}, 127 (2001).

\bibitem{PRD2002}
A.~E.~Motter and P.~S.~Letelier, Phys. Rev. D \textbf{65}, 068502 (2002).

\bibitem{CorFomSin:82} 
I.~Cornfeld, S.~Fomin, and Y.~Sinai, \emph{Ergodic Theory} (Springer, 1982). 

\bibitem{result0} The inverse transformation of (\ref{sua}) is well-defined since the integrability
of $r$ with respect to $\mu$ assures the integrability of $1/r$ with respect to $\mu_r$. 

\bibitem{local_global}
Along a {\it single} orbit, as considered in the transformation of Lyapunov exponents, the metric $\tilde{g}$ can always be extended as a single-valued function over the entire orbit; in periodic orbits, it can be extended as a single-valued function over the period of the orbit and as a multi-valued function beyond it.

\bibitem{longpaper}
This relation follows from the sub-exponential evolution 
of the angles between the Lyapunov 
vectors and holds for almost every $\boldsymbol x_o$.

\bibitem{grass1983}	
P.~Grassberger,  Phys. Lett. A \textbf{97}, 227 (1983).

\bibitem{hents1983}
H.~G.~E.~Hentschel and  I.~Procaccia, Physica D \textbf{8}, 435 (1983).

\bibitem{comment}
Similar definitions and results can be established for \textsf{limsup} replaced by \textsf{liminf}.

\bibitem{hyper}
This class of systems includes all geodesic flows on compact Riemannian manifolds with negative curvature 
\cite{BarRadWol:04}, which play a significant rule, for instance, in the characterization of chaos in
mixmaster cosmologies \cite{chitre72,burd93}.   

\bibitem{BarRadWol:04} 
L.~Barreira, L.~Radu, and C.~Wolf, Dyn. Syst. \textbf{19}, 89 (2004).

\bibitem{chitre72}
D.~M.~Chitre, Ph.D. Thesis, Univ. of Maryland (1972).

\bibitem{burd93}
A.~Burd and R.~Tavakol, Phys. Rev. D \textbf{47}, 5336 (1993).

\bibitem{entangled}
Similar behavior is found even when the two sets are intermingled. 
The details will be included in an extended paper.

\bibitem{ld1}
J.~L.~Kaplan and J.~A.~Yorke, Lect. Notes Math. \textbf{730}, 204 (1979).

\bibitem{ld2}
J.~D.~Farmer, E.~Ott, and J.~A.~Yorke, Physica D \textbf{7}, 153 (1983).


\end{thebibliography}
\end{document}